\documentclass[12pt]{article}

\normalbaselines
\begin{document}

\begin{center}

{\bf NON-MINIMAL ELECTRODYNAMICS}

\vspace{2mm}

{\bf AND RESONANCE INTERACTIONS IN RELATIVISTIC PLASMA}

\vspace{5mm}

{\bf A.B. Balakin \footnote{e-mail:Alexander.Balakin@ksu.ru} and
R.K. Muharlyamov  \footnote{e-mail:Ruslan.Muharlyamov@ksu.ru}}

\vspace{3mm}
{\it Kazan State University, Kremlevskaya str., 18,
420008, Kazan, Russia}

\end{center}

\vspace{1mm}

\begin{abstract}
A three-parameter toy-model, which describes a non-minimal
coupling of gravity field with electromagnetic field of a
relativistic two-component electrically neutral plasma, is
discussed. Resonance interactions between particles and
transversal waves in plasma are shown to take place due to the
curvature coupling effect.

\end{abstract}

\section{Introduction}

In order to explain the phenomenon of accelerated expansion of the
Universe and the nature of Dark Energy and Dark Matter numerous
modifications of the Einstein theory of gravity are elaborated,
the Non-minimal (NM) Field Theory being one of the most attractive
directions in these investigations. The NM Field Theory is based
on the introduction into the Lagrangian of the cross-invariants,
which contain the curvature tensor in convolutions with fields of
different nature and their covariant derivatives. The most
detailed variants of the NM theory are elaborated for the scalar,
electromagnetic and gauge fields (see, e.g., \cite{NM1,NM2}).

The basic principles of the NM Field Theory can also be used for
the NM modification of the relativistic kinetic theory of gas and
plasma \cite{gas1,NAbel}. Since the NM electrodynamics predicts
the tidal variations of the propagation velocity of the
electromagnetic waves, we expect that the NM theory of
relativistic plasma can exhibit new details of resonance
interactions of particles and plasma waves. In particular, it is
well-known that according to the minimal plasma theory the phase
velocity of transversal electromagnetic waves exceeds the speed of
light in vacuum, thus, there are neither resonance particle-wave
interaction nor Landau damping for such waves. In the NM extension
of plasma electrodynamics these phenomena are shown to be not
forbidden.

In this paper we consider a model, which can be indicated as a NM
modification of the well-known Vlasov theory \cite{Vlasov}. This
NM toy-model is based on a relativistic version of the Vlasov
kinetic equation \cite{Vlasov}, equations of plasma
electrodynamics \cite{LL} and NM extended Einstein equations
\cite{BL05}. For the illustration of the idea we use here a simple
toy-model assuming that the plasma and electromagnetic field are
the test ones, and the gravitational background is represented by
the de Sitter model.

\section{Master equations}

\noindent {\it (i)} {\it Relativistic kinetic equation}

\noindent In the context of the Vlasov theory a 8-dimensional
one-particle distribution function $f_{(a)}(x^i, p_k)$, which
describes particles of a sort "(a)" with the rest mass $m_{(a)}$
and electric charge $e_{(a)}$, and is a function of coordinates
$x^i$ and momentum four-covector $p_k$, is considered to satisfy
the relativistic kinetic equation of the following form
\cite{Vlasov}:
\begin{equation}\label{Vlasov}
p_l g^{il} \left[ \frac{\partial}{\partial x^i} + \Gamma^{s}_{ki}
p_s \frac{\partial}{\partial p_k}  + \frac{e_{(a)}}{c}F_{ki}
\frac{\partial}{\partial p_k} \right] f_{(a)} = 0 \,.
\end{equation}
Here $\Gamma^{s}_{ki}$ are the Christoffel symbols associated with
the space-time metric $g_{ik}$. The Maxwell tensor $F_{ki}$
describes a macroscopic averaged self-consistent electromagnetic
field in plasma, which guides the dynamics of charged particles.

\vspace{3mm} \noindent {\it(ii)} {\it Non-minimal electrodynamic
equations}

\noindent The master equations for the electromagnetic field have
the standard form:
\begin{equation}\label{Maxwell1}
\nabla_k H_{i}^{\ k} {=} {-} 4\pi \sum_{(a)} e_{(a)} N_i^{(a)} \,,
\end{equation}
\begin{equation}\label{Maxwell2}
 \nabla_i F_{kl} {+} \nabla_l F_{ik} {+} \nabla_k F_{li} {=} 0
\,,
\end{equation}
where $\nabla_k$ is the covariant derivative. According to the
Vlasov approach the particle number four-covector $N_i^{(a)}$ can
be represented by the following integral
\begin{equation}\label{current}
N_i^{(a)} {=} \int \frac{d_4 P}{\sqrt{{-}g}} \ p_i f_{(a)} \delta
\left[\sqrt{g^{ik}p_ip_k}{-}m_{(a)} c \right] \theta(V^kp_k)\,.
\end{equation}
Here $d_4P \equiv dp_0 dp_1 dp_2 dp_3 $ symbolizes the four-volume
in the momentum space; the delta function guarantees the
normalization property of the particle momentum, i.e.,
$g^{ik}p_ip_k = m^2_{(a)} c^2$; the Heaviside function $\theta(V^k
p_k)$ rejects negative energy, $V^k$ is a velocity four-vector of
the system as a whole. According to the NM approach
\cite{NM1,NM2,BL05}, the simplest form of the electromagnetic
excitation tensor $H^{ik}$ can be written as follows
\begin{equation}\label{excitation}
H^{ik} = F^{ik} + {\mathcal{R}^{ik}}_{mn} \ F^{mn} \,,
\end{equation}
where the NM susceptibility tensor
$$
{\mathcal{R}^{ikmn}} = \frac{q_1}{2}R \
(g^{im}g^{kn}-g^{in}g^{km}) +
$$
\begin{equation}\label{2}
\frac{q_2}{2} (R^{im}g^{kn} {-}R^{in}g^{km} {+}R^{kn}g^{im}
{-}R^{km}g^{in}) {+} q_3 R^{ikmn}
\end{equation}
contains three arbitrary parameters of NM coupling $q_1$, $q_2$
and $q_3$ with a dimensionality of length in square.

\vspace{3mm}

\noindent {\it (iii)} {\it Gravitational background}

\noindent The NM three-parameter extension of the Einstein
equations was discussed, e.g., in \cite{BL05}. When the plasma and
its electromagnetic field can be regarded as the test ones, one
can use the concept of background gravitational field. Below we
follow this concept and consider the de Sitter metric
\begin{equation}\label{metric}
ds^2= a^2(\eta)\left[d\eta^2-(dx^1)^2-(dx^2)^2-(dx^3)^2 \right]
\end{equation}
with $ a(\eta) = \frac{a_0}{\eta}$ as a gravitational background
for the NM plasma electrodynamics. For this metric the formulas
for the NM susceptibility (\ref{2}) yield
\begin{equation}\label{18}
{\mathcal{R}^{ik}}_{mn} = {\cal K} \ (\delta^i_m \delta^k_n -
\delta^i_n \delta^k_m)\,,
\end{equation}
where the constant $ {\cal K}$, defined as
\begin{equation}\label{K}
{\cal K} \equiv  - \frac{1}{a_0^{2}} \ (6q_1+3q_2+q_3) \,,
\end{equation}
may be positive, negative or equal to zero depending on the values
of NM parameters $q_1$, $q_2$ and $q_3$.

\section{Non-minimally coupled electromagnetic waves in a plasma}

Stationary 7-dimensional distribution function, the solution of
(\ref{Vlasov}), which guarantees that the macroscopic averaged
collective electromagnetic field in the electrically neutral
plasma vanishes, is well-known \cite{Bel}
\begin{equation}\label{Bel1}
f^{(\rm st)}_{(a)} = f^{(0)}_{(a)} (q^2) \,, \quad q^2 \equiv
p_1^2 {+} p_2^2 {+} p_3^2 \,,
\end{equation}
where $f^{(0)}_{(a)}(q^2)$ is arbitrary function of its argument.
The component $p_0$ of the particle momentum
\begin{equation}\label{Bel2}
p_0 = a^2 p^0 = \sqrt{m^2_{(a)}c^2 a^2(\eta) + q^2} \,,
\end{equation}
coincides with $q$ in the ultrarelativistic limit. The perturbed
quantities: distribution function $\delta f_{(a)}$ and the Maxwell
tensor $\delta F_{ik} \equiv F_{ik}$, satisfy in our toy-model to
the system of integro-differential equations:
\begin{equation} \label{toy1}
\left[q \ \frac{\partial }{\partial \eta} + q^\gamma
\frac{\partial}{\partial x^\gamma} \right] \delta f_{(a)} =
\frac{e_{(a)}}{c} \  F_{\gamma 0} \ q^\gamma \ \frac{\partial
f^{(0)}_{(a)}}{\partial q} \,,
\end{equation}
\begin{equation} \label{toy2}
(1{+}2{\cal K}) \frac{\partial}{\partial x^\alpha} F_{0\alpha} =
4\pi \sum_{(a)} e_{(a)} \int d^3q  \ \delta f_{(a)} \,,
\end{equation}
$$ (1+ 2{\cal K})
\left[\frac{\partial}{\partial x^\alpha}F_{\gamma\alpha} -
\frac{\partial}{\partial \eta} F_{\gamma 0}\right] =
$$
\begin{equation} \label{toy3}
\ \ \  = - 4\pi \sum_{(a)} e_{(a)} \int \frac{q^{\gamma}}{q} d^3q
\ \delta f_{(a)} \,,
\end{equation}
\begin{equation} \label{toy4}
\frac{\partial F_{\gamma\alpha}}{\partial \eta}+\frac{\partial
F_{\alpha 0}}{\partial x^\gamma}+\frac{\partial F_{0
\gamma}}{\partial x^\alpha} = 0 \,.
\end{equation}
Here we introduced convenient notations
\begin{equation}
\label{15} q^\alpha \equiv a^2 g^{\alpha \gamma} p_{\gamma} \,,
\quad q^2 {=} (q^1)^2 {+} (q^2)^2 {+} (q^3)^2 \,,
\end{equation}
and $d^3q \equiv dq^1 dq^2 dq^3$, the summation over the repeating
indices being assumed. The non-minimal equations (\ref{toy1}) -
(\ref{toy4}) form a system of linear integro-differential
equations with coefficients, which do not depend on time, and it
is a good surprise. This system can be resolved using the Fourier
transformation
\begin{equation} \label{Fourier}
{\bf H}(\eta, x^{\alpha}) {=} \int_{{-}\infty}^{\infty}
\frac{d\omega d_3k}{(2\pi)^4} {\bf {\cal H}}(\omega,
k_{\alpha})e^{i\left( k_{\alpha} x^{\alpha} {-} \frac{\omega
\eta}{c}\right)} \,,
\end{equation}
where calligraphic letter  ${\bf {\cal H}}(\omega, k_{\alpha})$
denotes the Fourier transform of the function ${\bf H}(\eta,
x^{\alpha})$; $d_3k \equiv dk_1dk_2dk_3$. Simple calculations give
\begin{equation}\label{F1}
\delta f_{(a)} (\omega, k_{\alpha}) {=} {\cal F}_{\gamma
0}(\omega, k_{\alpha}) \left\{ \frac{ie_{(a)} q^\gamma}{q
\left(\omega {-} \frac{c k_{\alpha} q^{\alpha}}{q}\right) } \
\frac{\partial f^{(0)}_{(a)}}{\partial q}\right\} \,,
\end{equation}
\begin{equation}\label{basic1}
{\cal F}_{\alpha 0} \left[(1{+}2{\cal K}) \frac{k^2c^2}{\omega^2}
\left(\frac{k_{\alpha} k_{\beta}}{k^2} {-} \delta_{\alpha \beta}
\right) {+} 2 {\cal K} \delta_{\alpha \beta} {+}
\varepsilon_{\alpha \beta} \right] {=} 0 \,,
\end{equation}
where $\varepsilon_{\alpha \beta}$ denotes a permittivity
three-tensor
\begin{equation}\label{permitt}
\varepsilon_{\alpha \beta} \equiv \delta_{\alpha \beta} {+}
\frac{4\pi}{\omega} \sum_{(a)} e^2_{(a)} \int \frac{ d^3q \
q^\alpha q^\beta}{q^2 \left(\omega {-}
\frac{ck_{\gamma}q^{\gamma}}{q} \right)} \ \frac{\partial
f^{(0)}_{(a)}}{\partial q} \,.
\end{equation}
Since the zero-order distribution function $f^{(0)}_{(a)}$ is
considered to depend on $q^2$, the permittivity tensor has an
isotropic structure
\begin{equation}\label{ISO}
\varepsilon_{\alpha\beta} = \left(\delta_{\alpha\beta} -
\frac{k_\alpha k_\beta}{k^2}\right) \varepsilon^{({\rm tr})} +
\frac{k_\alpha k_\beta}{k^2} \varepsilon^{({\rm l})} \,,
\end{equation}
thus, the dispersion relations for the longitudinal and
transversal electromagnetic waves are, respectively,
\begin{equation}\label{DE1}
1 - \varepsilon^{({\rm l})}(\omega, k_{\alpha}) = 1 + 2 {\cal K}
\,,
\end{equation}
\begin{equation}\label{DE2}
1- \varepsilon^{({\rm tr})}(\omega, k_{\alpha}) = (1+2 {\cal
K})\left(1-\frac{k^2c^2}{\omega^2}\right) \,.
\end{equation}
When ${\cal K}= 0$, the equations (\ref{DE1}) and (\ref{DE2})
coincide with the well-known dispersion relations (see, e.g.,
\cite{LL}), as it should be. Since the scalar permittivities
$\varepsilon^{({\rm l})}$ and $\varepsilon^{({\rm tr})}$ do not
depend on the NM parameters, they coincide with the ones, obtained
by Silin in the framework of  minimal theory of ultrarelativistic
plasma \cite{Silin}.

The dispersion relations for longitudinal plasma waves (see
(\ref{DE1})) and for transversal electromagnetic waves (\ref{DE2})
depend essentially on the sign of the parameter $1 {+} 2 {\cal K}$
with ${\cal K}$ given by (\ref{K}). This parameter is
predetermined both by the space-time curvature and by constants of
non-minimal coupling. Below we consider all three cases, when this
parameter is positive, negative or vanishes.

\noindent (I) {\it First case: $1 {+} 2 {\cal K} > 0$}

\noindent Using the NM reparametrization of the Debye radius
$r_{{\rm D}} \to r^{*}_{{\rm D}} \equiv r_{{\rm D}} \ \sqrt{1 {+}
2 {\cal K}}$, where
\begin{equation}
\frac{1}{r^2_{{\rm D}}} \equiv \sum_{(a)} \frac{4\pi e^2_{(a)}
N_{(a)}}{k_B T_{(a)}} \,, \label{Debye}
\end{equation}
(see \cite{LL,Silin}), one can show explicitly, that all the
results for the longitudinal and transversal waves \cite{LL,Silin}
in the ultrarelativistic plasma remain (qualitatively) valid.

\noindent (II) {\it Second case: $1 {+} 2 {\cal K} = 0$}

\noindent Such a value of the parameter ${\cal K}$ relates to the
case of vanishing excitation tensor $H^{ik}$. The Maxwell
equations (\ref{Maxwell1}) remain consistent if the electric
current vanishes. This means that self-consistent averaged
electromagnetic field should also vanish (see (\ref{toy2}),
(\ref{toy3}) and (\ref{F1})), thus, there are no plasma waves in
such a system.

\noindent (III) {\it Third case:$1 {+} 2 {\cal K} < 0$}

\noindent This case gives principally new results. After the NM
reparametrization of the Debye radius $r_{{\rm D}} \to r^{*}_{{\rm
D}} \equiv r_{{\rm D}} \ \sqrt{|1 {+} 2 {\cal K}|}$ in terms of
dimensionless variables
\begin{equation}
z = x + i y  \equiv \frac{\omega}{kc} = \frac{\Omega}{kc} + i
\frac{\gamma}{kc} \,, \label{TRANS0}
\end{equation}
where $\Omega(k)$ is a frequency of plasma oscillations, and
$\gamma(k)$ is a decrement of damping or increment of instability
(depending on its sign), the dispersion relation for {\it
longitudinal} plasma waves takes the form
\begin{equation}
2\left(k^2 r^{*2}_{{\rm D}} -1 \right) = \Phi(z) \,,
\label{Silin11}
\end{equation}
where
\begin{equation}
\Phi(z) \equiv  z \log{\frac{|z-1|}{|z+1|}} + i \pi z \Theta
\left(1{-}\Re e z \right)\,. \label{S11}
\end{equation}
In the minimal plasma electrodynamics there exists a solution with
$k=0$ and $\omega = c / \sqrt{3}r_{{\rm D}}$, but the solution
with $\omega=0$ for real wave numbers $k$ does not exist. The
situation in the NM plasma electrodynamics can be just opposite:
the equation (\ref{Silin11}) admits the solution $\omega = 0$,
when the wave length is equal to the modified Debye radius, i.e.,
$k r^{*}_{{\rm D}} = 1$. As for a solution with $k \to 0$, the
equation (\ref{Silin11}) gives $\omega^2 + c^2 / 3r^{*2}_{{\rm
D}}=0 $, i.e., there are no oscillations, the process of plasma
evolution is aperiodic.

When we deal with {\it transversal} electromagnetic waves, the
dispersion relation reduces to
\begin{equation}
\frac{2z^2}{1-z^2} - 4k^2  r^{*2}_{{\rm D}} = \Phi(z)  \,.
\label{TRANS1}
\end{equation}
Numerical analysis of the solution $z(k)$ of this equation (we
omit it in this short report) shows that there exist solutions
with $\Re e z < 1$. This principally new result can be illustrated
qualitatively in the case, when the imaginary part of the solution
$z(k)$ is much smaller than the real one, i.e., $|\gamma| <<
|\Omega|$. Then the real part $x=\Re e z$ should be found from the
real equation
\begin{equation}
4k^2  r^{*2}_{{\rm D}} = \frac{2x^2}{1-x^2} - x
\log{\frac{|x-1|}{|x+1|}} \equiv \Psi(x) \,. \label{S12}
\end{equation}
Clearly, the function $\Psi(x)$ increases monotonically from zero
at $x=0$ to infinity, when $x \to 1_{-0}$, remaining positive for
arbitrary $x$ from the interval $0<x<1$. Thus, for each value of
positive parameter $4k^2 r^{*2}_{{\rm D}}$ one can find the
corresponding point on the plot of $\Psi(x)$, so, the solution
$x=X(k)$ exists for all $k$. Analogously, it is easy to check,
that for the interval $x>1$ there are no solutions. Since the
dispersion relation admits the solution with the phase velocity
less that speed of light in vacuum ($\Omega /k < c$), the
imaginary part of frequency, $\gamma$, is non-vanishing:
\begin{equation}
\gamma = \frac{1}{4}\pi kc \
\frac{\left(\frac{\Omega}{kc}\right)^2
\left[1-\left(\frac{\Omega}{kc}\right)^2
\right]^2}{\left(\frac{\Omega}{kc}\right)^2 + \left[1 -
\left(\frac{\Omega}{kc}\right)^2 \right]^2 k^2 r^{*2}_{{\rm D}}}
\,.
\label{TRANS4}
\end{equation}
When $1{+}2{\cal K}$ is negative, the parameter $\gamma$ becomes
positive, i.e., it describes an increment of instability. The
parameter $\gamma$ disappears, when $\Omega {=} 0$ and when
$\Omega {=} kc$, thus, being positive, it reaches a maximum on the
interval $0<\Omega<kc$. For small values of $\Omega / kc$ the
condition $\gamma << \Omega$ is valid, when $k^2 r^{*2}_{{\rm D}}
>> 1$; for $\Omega \simeq kc$, it is valid for arbitrary value of
the parameter $kr^{*}_{{\rm D}}$.

\section{Conclusions}

\noindent 1. The discussed non-minimal toy-model demonstrates an
explicit example that the effect of curvature coupling provides
transversal electromagnetic waves to propagate in a relativistic
plasma with a phase velocity less than speed of light in vacuum.

\noindent 2. Curvature coupling can support resonance interactions
of transverse electromagnetic waves with plasma particles.

\noindent 3. In case of domination of the non-minimal interactions
($2{\cal K}<-1$) the resonance interactions of plasma particles
with transversal electromagnetic waves produce an instability,
which is an antipode of Landau damping.

\vspace{1mm} \noindent{\bf Acknowledgments}

\noindent A.B. thanks DFG (project No. 436RUS113/487/0-5). This
work was partially supported by RFBR (grant N 08-02-00325-a, and
grant N 06 -01-00765).

\end{document}